# Phase diagram of the sulfur hydride: transition into high $T_c$ state

Lev P. Gor'kov[1] and Vladimir Z. Kresin[2]*

[1] *NHMFL, Florida State University, Tallahassee, Florida, 32310, USA*

[2] *Lawrence Berkeley Laboratory, University of California, Berkeley, CA 94720, USA*

Remarkable feature of the phase diagram of the record high $T_c$ superconductors, sulfur hydrides, is a sharp increase from $T_c \approx 120K$ up to $T_c \approx 200K$. This increase is a signature of the structural transition. The study described below is concerned with the nature of this phase transition. One can demonstrate that the symmetry analysis along with an analysis of the impact of lattice deformations lead to the conclusion that we are dealing with the first order transition. It leads to an abrupt appearance of small pockets on the Fermi surface and, correspondingly, to the two-gap energy spectrum.

This paper is concerned with the phase diagram, that is, with the dependence of the critical temperature on pressure for the sulfur hydrides. This material displays the record value of $T_c \approx 203K(!)$ [1], [2]. The most remarkable feature of the phase diagram is that the value of $T_C \approx 120K$ at $P \approx 125 GPa$ sharply increases to $T_C \approx 200K$ at $P \approx 150 GPa$ as if in the 1$^{st}$ order phase transition (Fig.1)

The problem of the nature of the transition has attracted a lot of interest. Identifying the crystalline symmetry of these phases can essentially help to solve this problem. The theoretical calculations [3]-[7] suggest that the high - $T_c$ phase at pressures $P \geq 150 GPa$ has the body-centered cubic symmetry of the cubic group Im-$3m$). The result is consistent with the X-rays data [2],[8].

In our paper [9] we have assumed that the sharp transition into the high $T_c$ state is of the first order. It allowed us toexplain the slow decrease in $T_c$ at pressures P>$P_{cr}$ ($P_{cr}$ corresponds to $T_{c;max}$=203K). The present paper contains a more detailed analysis of this issue.

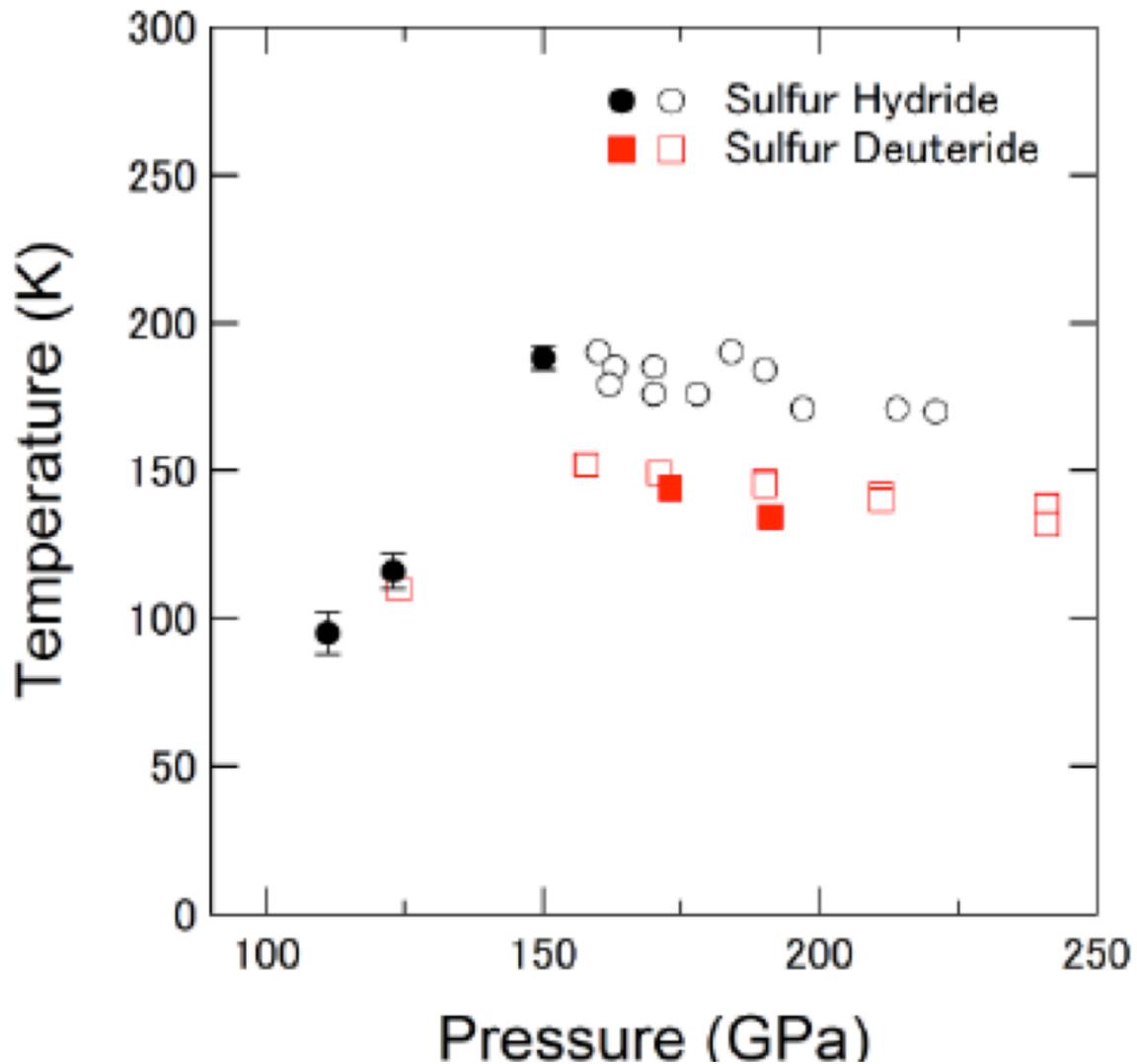

Fig.1. Pressure dependence of $T_c$. The data for annealed samples are presented. One can see a large increase in the value of $T_c$ in the region near P=150GPa. Adapted from Einaga et al., 2016.

Currently, the consensus in literature is that the Bravais lattices of the high-$T_c$ and the low-$T_c$ phases belong to different symmetries (the *Im-3m* group for the cubic phase and the trigonal *R3m* symmetry group for the low-$T_c$ phase) and that the structural phase transition between them occurs at a pressure somewhere in between $P \leq 150 GPa$ and $P \geq 125 GPa$.

## 1. Structural transition and sharp increase in $T_c$

The observed rapid variation of $T_c$ in the given pressure interval immediately poses the question regarding the character of the phase transition between these two phases. In [9] we have assumed that the near doubling of $T_c$ in the narrow experimental pressure interval $\Delta P \approx 25 GPa$ is the signature of a first-order structural phase transition between the phases with the lower and higher $T_c$. The data plotted in Fig. 1, obtained both while increasing and decreasing the pressure point out at the discontinuous character of the transition. However, character of a structural transition cannot be deduced unambiguously only from the pressure dependence of $T_c$.

The problem should be approached first in frameworks of the Landau theory of the symmetry phase transitions [10] Assume that in a vicinity of the point of the transition $T_c$ the thermodynamic potential of the high-symmetry phase $\Phi(T,P)$ is expanded in the series of a parameter $\eta$ with a lower crystalline symmetry:

$$\Phi = \Phi_0 + A\eta^2 + B\eta^3 + C\eta^4 + ... \qquad (1)$$

By taking the first term in the form $A(T) = a(T - T_C)$ one points out that the system becomes unstable at $T < T_C$. The necessary requirement for the second order phase transition is the absence of the third order term $B(T_C) \equiv 0$, so that the thermodynamic potential can have a minimum at the transition point. If $B(T_C) \neq 0$, the phase transition is of the first order and is accompanied by a sudden change (jump) in the lattice structure. Whether the coefficient $B(T_C)$ equals to zero or not depends on the symmetry of the order parameter $\eta$. In particular, in our case of the cubic group *Im-3m* ($O_h$) it is the question whether the order parameter $\eta$ is invariant with respect of the spatial inversion. (In that follows, we find more convenient to discuss the phase transformation between phases with the *Im-3m* and the *R3m* symmetry as a function of decreasing pressure).

The cubic space group *Im-3m* ($O_h$) contains the *inversion* among its symmetry elements, while in the space group #160 (*R3m*) belonging to the class $C_{3v}$ for which the *inversion* is absent. Hence, the second-order structural transition between the high- $T_c$ *Im-3m* phase and the phase *R3m* does not contradict to the Landau theory.

The structural transition into the lower symmetry state without a change in the lattice periodicity is driven by softening of a vibration mode at the center of the Brillouin Zone (BZ). Thus, according to [6],[7], the transition between the *Im-3m* and the *R3m* phases is driven by the sulfur-hydrogen stretching mode. One can assume that this specific result, may be sensitive to details of the calculation. Indeed, in [7] one finds $P_{cr} = 150 GPa$ for the critical pressure $P_{cr}$, whereas $P_{cr} = 103 GPa$ in [6].

We have performed the group-theoretical symmetry analysis (see Appendix), and have shown *rigorously* that the list of the phonon modes at the center of the Brilloiun Zone (BZ) for the point group $O_h = O \times C_i$ is comprised of the four *odd* three-dimensional irreducible representations (the three vector representations $F_{1u}$ and one $F_{2u}$, see [11]), so that *every* phonon instability with the zero structural vector $\vec{Q} = 0$ would lead to the phase transition of the second order.

Nevertheless, the mechanisms governing the transition between the two phases can be of a more complicated origin. The phonon softening can be caused by the electron-phonon interaction and the corresponding renormalization of the phonon frequency (see, e.g., [12],[13] ).In the general case, one must keep in mind the possibility of the phonon softening at the momentum away from the center of the BZ.

The "imaginary phonon frequencies" appear in $H_3S$ (in the harmonic approximation) at several points of the BZ [6].The instability in this case would correspond to the developing below $T_c$ of a lattice deformation with a non-zero structural vector $\vec{Q} \neq 0$. Such possibility, for instance, realizes in the charge density wave (CDW) transitions (see [14]).

The CDW instability with a non-zero structural vector has been studied for the transition-metal dichalcogenides[15]).The incommensurate CDW phase that develops directly below the instability point gives place to the commensurate CDW in the *first-order* phase transition. The *trigonal R3m* phase with three $H_3S$ entities per unit cell drawn in Fig. 2 ,Ref. [3]. could serve as the illustrative example of such commensurate modulated phase in H3S.

Again, to the best of the author's knowledge, softening of a phonon frequency $\omega(\vec{Q})$ owing renormalization of the frequency by the electron-phonon interactions has never been discussed in the hydrides.

Although the X-rays measurements [2] provided an important information regarding symmetry of the $H_3S$ lattice, these data describe with sufficient accuracy only the positions of the S-ions, but not light H-ions. One may expect that the future X-rays measurements with the higher resolution will provide an additional information about changes in the crystalline structures.

According to several theoretical prediction, the Fermi surface of the high-$T_C$ phase may contain small pockets, which appear simultaneously with the onset of the high-$T_C$ phase itself. The idea of the first order transition allows us to explain self-consistently the slow decrease in $T_C$ with an increase in pressure above the pressure $P_{cr}$ corresponding to the maximum of $T_C = 203K$ [1],[2].

Let us discuss if there are mechanisms that would allow interpretation of the observed rapid variation of $T_C$ as a function of pressure in terms of a first order transition. The above postulate of the symmetry second order transition ceases to be correct if the interactions between the order parameter and the lattice deformations (that is, the so-called striction effects) were added into the consideration. (see the next section).

## 2. Quadratic striction and a first order transition.

Let us discuss here the first-order transition obliged to the interaction between the internal degrees of freedom that govern the Landau second order symmetry transition and the acoustic lattice deformations. The structural vector remains at the center of the Brilloiun Zone. In the extensive theoretical literature on high $T_C$ superconductivity in sulfur hydride H$_3$S the role of the so-called quadratic striction has not been explored yet.

Coupling to the lattice can transform a second order transition into the weak first-order transition accompanied by the structural changes. This result has been rigorously proven for the elastically isotropic solid in [16] by taking the fluctuations into account. Although the method developed by them does not apply to the anisotropic materials, the solution undoubtedly reflects the genuine physics in the general case as well. Therefore, to study the effect of the quadratic striction on the displacive phase transitions for the cubic *Im-3m* phase, we apply the simplified approach (see also [17])

As the order parameter let us choose the displacements $\vec{S}_h$ of a hydrogen atom from its equilibrium position in the middle between the sulfurs in the *Im-3m* phase (the so-called stretching mode; the vector representations is $F_{1u}$). The Landau functional near the temperature of transition $T_0$ is of the form:

$$\Omega(T - T_0) = \int \frac{a}{2}\{\tau \vec{S}_h^2 + b(\vec{S}_h^2)^2 + c(\xi_0 \nabla \vec{S}_h)^2\} dr \qquad (2)$$

Here $\tau = T - T_0$, $\xi_0$ is a coherence length and *a*, *b* and *c* are some material constants.

Add to the functional (2) all linear in the deformation expressions for the interaction between $\vec{S}_h$ and the lattice. Not to overcrowd the formulas, we present the latter schematically in a form:

$$H_{str} = -q \int \hat{u}(r) \vec{\hat{S}}_h^2(r) d\vec{r} . \qquad (3)$$

Here the notation $\hat{u}(r)$ stands for the components of the strain tensor:

$$u_{ik} = \frac{1}{2}\left(\frac{\partial u_i}{\partial x_k} + \frac{\partial u_k}{\partial x_i}\right).$$

We write down the elastic energy also in the same schematic form:

$$H_{el} = \int \hat{K} \hat{u}(r)^2 d\vec{r} \qquad (4)$$

In Eq.(4) $\hat{K}$ stands for the various elastic moduli in front of the corresponding combination quadratic in the strain tensor $\hat{u}(r)$.

The Gibbs energy $\Phi(T)$ near the temperature of the phase transition is:

$$\Phi(T) = \Phi(T_0) + \Omega(T - T_0) + H_{str} + H_{el} \qquad (5)$$

The second term $H_{str}$ (see Eq. (3)), linear in the deformation and quadratic in the order parameter, provides the formal definition of the so-called quadratic striction. Its role consists first of all in the following formal substitution in Eq. (2):

$$\tau \to \tau + \lambda \hat{u} \qquad (6)$$

(For example, in case of coupling in (6.3) with a single elastic mode $\lambda = -2q/a$). After substituting $\tau \to \tau + \lambda \hat{u}$ Eq. (5) acquires the form:

$$\Phi(T) = \Phi(T_0) + \Omega(\tau + \lambda \hat{u}) + \int \hat{K}\hat{u}(r)^2 \, d\vec{r} \qquad (7)$$

Let $\hat{u}_{ext}$ be a *homogeneous* external deformation $\tau \to \tau + \lambda \hat{u}_{ext}$. Differentiating $\Omega(\tau + \lambda \hat{u}_{ext})$ below $T_0$, one finds:

$$\frac{\partial^2 \Phi}{\partial \hat{u}_{ext}^2} = \lambda^2 \frac{\partial^2 \Phi}{\partial \tau^2} = -\frac{\lambda^2 C(\tau)}{T_0} \qquad (8)$$

where $C(\tau)$ is the specific heat.

That is, in the Landau theory the jump at the transition $\Delta C$ in the specific heat is accompanied by a step-like negative change (8) in the elastic moduli. As $\Delta C \propto T_0$, the change in a modulus is of the order of the modulus itself.

Outside the fluctuation temperature interval all corrections from Eq. (8) are finite. Inside the fluctuation regime $|\Delta T/T_0| \ll 1$ the specific heat has a singularity $C(\tau) \propto |\tau|^{-\alpha}$ (see [16]). Formally, according to Eq. (8), the elastic moduli will diverge at the temperature of transition and the lattice becomes unstable. Fortunately, there is no need to come into the details of the behavior of all variables so close to the singularity point. The latter is made irrelevant by the first order transition in the system that owes its origin to the fluctuations inside a finite temperature interval, i. e., far from the singularity [16] The transition changes its character from the second order to that of the weak first order.

For the purpose of the demonstration limit ourselves by the first singular correction to the Gibbs potential (see [10], §146)

$$\Omega_{sing}(\tau) = -B|\tau|^{3/2} \tag{9}$$

and consider coupling in (3) with one elastic mode $u$ only:

$$\Phi_{sing}(\tau;u) = \Omega_{sing}(\tau + \lambda u) + Ku^2 \tag{10}$$

Minimizing $\Phi_{sing}(\tau;u)$ with respect to $u$ ($\partial\Phi_{sing}(\tau;u)/\partial u = 0$) obtain:

$$\partial\Omega_{sing}(x)/\partial x = -2(K/\lambda)u, \tag{11}$$

where $x = \tau + \lambda u$. From Eq. (11)

$$\lambda u = \lambda^2 B \frac{3x}{4K|x|}|x|^{1/2}. \tag{12}$$

Eq. (12) together with Eqs. (6) and (9) determine the thermodynamics of the transition.

To start with, consider the relation between $x$ and the temperature $\tau$:

$$x = \tau + \lambda^2 B \frac{3x}{4K|x|}|x|^{1/2}. \tag{13}$$

At small $\tau$ Eq. (13) has three solutions for $x$. Introducing the parameter $a = 3\lambda^2 B/4K$ rewrite Eq. (13) in the new variables $x = a^2 \bar{x}$ and $\tau = a^2 \bar{\tau}$:

$$\bar{\tau} = \bar{x} - \frac{\bar{x}}{|\bar{x}|}|\bar{x}|^{1/2} \qquad (14)$$

The dependence $\bar{\tau}(\bar{x})$, Eq.(14) is shown in Fig.2

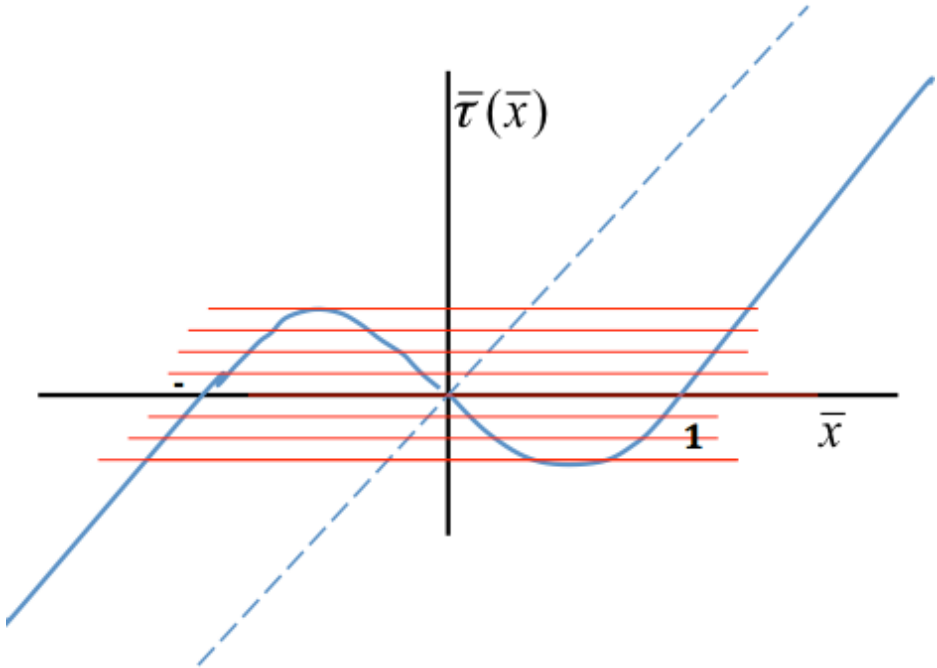

Fig.2. The hysteresis describing the weak first order transition accompanied by the structural changes, Eq.(6.14). Here $\bar{x} = a^2 x$, $\bar{\tau} = a^2 \tau$, $a = const, x = \tau + \lambda u, \tau = T - T_0$, $\lambda$ is the elastic constant, u is the elastic mode, $T_0$ is the transition temperature.

At small enough $\bar{\tau}$ Eq.(14) has three solutions for $\bar{x}$, which is the characteristic feature of the 1$^{st}$ order transition.

According to Eq. (9), $\Omega_{sing}(x) = -B|x|^{3/2}$. Once the two branches $x_+(\tau)$ and $x_-(\tau)$ are chosen, the point of the transition is to be found from the equation $\Omega_{sing}(x_+) = \Omega_{sing}(x_-)$. As one can see directly from Fig.5 the transition takes place at $\bar{\tau} = 0$.

For the applicability of this analysis the transition must involve into the consideration fluctuations in a narrow but finite temperature interval $|\Delta T/T_0|<<1$ where the specific heat has a singularity $C(\tau) \propto |\tau|^{-\alpha}$. For our choice of $\alpha = 1/2$ the definition of the dimensionless variables in Eq. (14) implies that $a^2 = (3\lambda^2 B/4K)^2 << T_0$.

Extension of the above arguments on the case of the phase transition driven by the applied pressure is possible. The pressure interval [1],[2] al., where $T_c$ rapidly varies is narrow: $\Delta P/P \approx 1/5$. This is consistent with the ideas of the striction model. We therefore suppose that the accuracy of the *ab initio* calculations (see e.g. [4]) was not sufficient to resolve the step-like character of the transition between the phases with lower and higher $T_c$.

**Appendix.**
**Classification of the normal vibrations of H$_3$S at the $\Gamma$-point \\ ($\Omega(\vec{k}=0)$).**

a) The total symmetry of *H$_3$S* is given by the cubic group $O_h = O \times C_i$, where $O$ is the group of all the symmetry rotations of a cube and $C_i$ is the inversion group.

b) For classifying the normal vibrations of a crystal at $\vec{k}=0$ it is necessary to find the complete vibrational representation realized by all vibrational coordinates (see, e.g. ,[16],§ 136). The latter in the case of $H_3S$ is comprised of the atomic displacement of sulfur and of the three hydrogen ions. That is, one must find the twelve -dimensional representation built on the variables of four polar vectors.

c) Recall several main results from the theory of the point groups (see [17]). More specifically, enumerate the symmetry transformation and characters of the irreducible representations of the cubic groups $O$ and $O_h = O \times C_i$.

Elements of the group are distributed over the classes. (Thus, the two rotations with respect to the threefold axis $C_3$ and $C_3^2$ fall in one class). As characters are the same for the elements of a class, everywhere below are given the sum of characters of a class.

Secondly, as the group $O_h$ is the product of the group $O$ and the inversion, all the irreducible representations of the group $O_h$ are comprised of the even (g) and the odd (u) representations of the former.

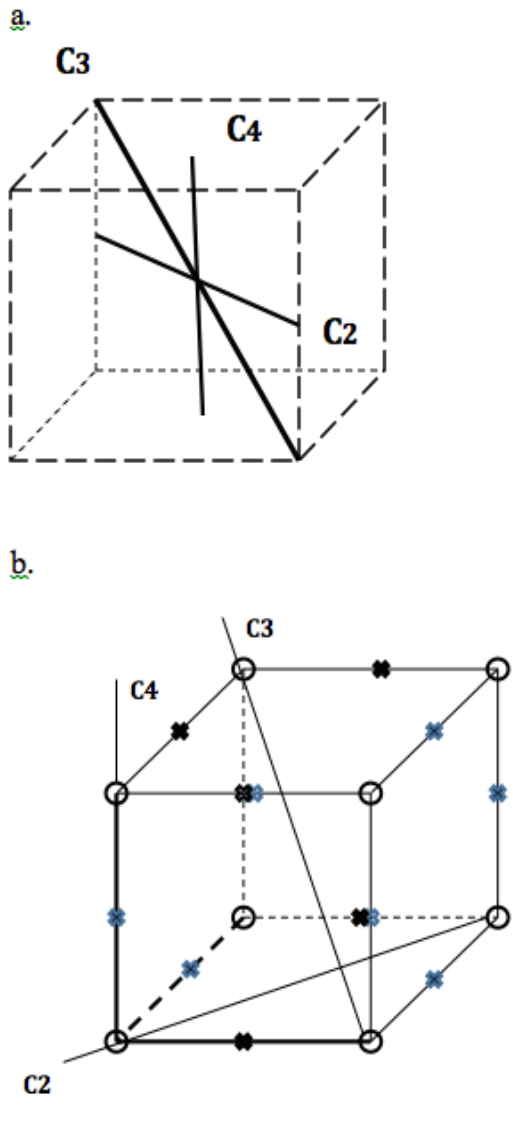

Fig.3. Rotational symmetry of cubic $H_3S$: (a) axes of the rotational symmetry of the cube.; (b) $H_3S$: the unit cell .The bold lines single out the basic periods of the lattice; the empty circles stand for the sulfur atoms, while the crosses mark positions of the hydrogen atoms. Three rotational axes from Fig.12a are shown schematically.

Leaping ahead, we shall deal below only with the threefold representations $F_{1u}$ and $F_{2u}$ (see [17]). The table below enumerate characters of both representations:

Table 1. Characters of the representations

| $\hat{e}$ | $E$ | $8C_3$ | $6C_4$ | $3C_2$ | $6C_2$ | $I$ | $8S_3$ | $6S_4$ | $3\sigma_h$ | $6\sigma_v$ |
|---|---|---|---|---|---|---|---|---|---|---|
| $F_{1u}$ | 3 | 0 | 1 | −1 | −1 | −3 | 0 | −1 | 1 | 1 |
| $F_{2u}$ | 3 | 0 | −1 | −1 | 1 | −3 | 0 | 1 | 1 | −1 |

(In the notations in Table 1 it was used that $C_n \times I = S_n$ and $C_3 \times I = \sigma_{h(v)}$

e) The rules for composing characters of the complete vibrational representation [16] are based on the self evident fact that only the atoms that do not move or go over into the equivalent lattice positions at a given transformation can contribute into these characters.

Let $v_C$ be the number of such atoms for a transformation $C$ from the group $O$; correspondingly, $v_S$ is the number of such atoms for a given transformation $S_n$. Their contributions $\lambda_{C(S)}$ into the characters of the complete vibrational representation are obtained by multiplying $v_C$ and $v_S$ by the character of an irreducible representation of the polar vector $\lambda_{C(S)} = v_{C(S)} + \chi_{C(S)}$

where

$$\chi_C = 1 + 2\cos(\varphi_C) \qquad (A.1)$$

and

$$\chi_S = 1 + 2\cos(\varphi_S) \qquad (A.2)$$

(Standing for the reflection in a plane is $S(0)$; for the inversion in a center of symmetry $S(\pi)$).

d) Consider, for illustration rotation $C_4$. There are the two atoms (sulfur and one hydrogen on the fourfold axis per period) that do not shift under the rotation by the angle $\varphi_c = \pi/2$. With $\nu_c = 2$ and $\chi_{C_4} = 1$ one finds $\lambda_{C_4} = 2$.

After a tedious analysis of the each transformation listed in Table 1 we find:

Table 2. Vibrational representation

| $\hat{e}$ | $E$ | $8C_3$ | $6C_4$ | $3C_2$ | $6C_2$ | $I$ | $8S_3$ | $6S_4$ | $3\sigma_h$ | $6\sigma_v$ |
|---|---|---|---|---|---|---|---|---|---|---|
| $\lambda_{C(S)}$ | 12 | −1 | 2 | −4 | −2 | −12 | 1 | −2 | 4 | 2 |

With the use of characters in Table 1 for the threefold representations $F_{1u}$ and $F_{2u}$ it is now straightforward to show that the complete vibrational representation (Table 2) decomposes into the sum $F_{2u} + 3F_{1u}$.

We are grateful to M.Calandra and M.Eremets for valuable discussions. The work of LPG is supported by the National High Magnetic Field Laboratory through NSF Grant No. DMR-1157490,the State of Florida and the U.S. Department of Energy. The work of VZK is supported by the Lawrence Berkeley National Laboratory, University of California at Berkeley, and the U.S. Department of Energy.

* Corresponding author: vzkresin@lbl.gov